\documentstyle[times,graphics,astrobib,amssymb,epsfig]{mn2e}

\def\etal{et al.\ }
\def\ie{i.\,e.\,, }
\def\eg{e.\,g.\,}

\def\unit #1{\,{\rm #1}}

\def\arcmin{\unit{arcmin}}

\def\re{r_{\rm e}}
\def\bba{\left<}
\def\eba{\right>}
\def\mean #1{\bba #1 \eba}
\def\mubre{\mu_{\rm b}(<\re)}
\def\mmubre{\mean{\mubre}}

\def\gne #1#2{\ \vphantom{S}^{\raise-0.5pt\hbox{$\scriptstyle#1$}}_
{\raise0.5pt \hbox{$\scriptstyle#2$}}}

\def\labequn #1{\label{eq:#1}}

\def\araa#1{{\it Ann. Rev. Astron. Astrophys. }{\bf #1}}

\def\mean #1{\bba #1 \eba}
\def\unit #1{\,{\rm #1}} 

\def\re{r_{\rm e}}
\def\ibzero{I_{\rm b}(0)}

\def\rd{r_{\rm d}}

\def\idzero{I_{\rm d}(0)}

\def\mubzero{\mu_{\rm b}(0)}
\def\mubre{\mu_{\rm b}(<\re)}
\def\mmubre{\mean{\mubre}}

\def\mudzero{\mu_{\rm d}(0)}

\title[NIR bulge-disk correlations for lenticulars]
{Near-Infrared bulge-disk correlations of lenticular galaxies}
\author[Barway \etal]
{Sudhanshu Barway $^{1,\ 3}$\thanks{E-mail: barway@saao.ac.za},
Yogesh Wadadekar $^{2}$, Ajit K. Kembhavi $^{3}$, and Y. D. Mayya $^{4}$\\
$^{1}$South African Astronomical Observatory, P.O. Box 9, 7935, Observatory, Cape Town, South Africa; \\ 
$^{2}$National Centre for Radio Astrophysics, Post Bag 3, Ganeshkhind, Pune 411007, India; \\ 
$^{3}$Inter University Centre for Astronomy and Astrophysics, Post Bag 4, Ganeshkhind, Pune 411 007, India; \\ 
$^{4}$Instituto Nacional de Astrofisica, Optica y Electronica, Luis Enrique Erro 1, Tonantzintla, Apdo Postal 51 y 216,
C.P. 72000, Puebla, M\'exico \\ } 

\begin{document}

\maketitle


\begin{abstract}

We consider the luminosity  and environmental dependence of structural
parameters  of  lenticular galaxies  in  the  near-infrared $K$  band.
Using  a  two-dimensional  galaxy  image decomposition  technique,  we
extract  bulge and  disk  structural  parameters for  a  sample of  36
lenticular galaxies observed by us  in the $K$ band. By combining data
from the literature  for field and cluster lenticulars  with our data,
we study  correlations between parameters that  characterise the bulge
and the  disk as  a function of  luminosity and environment.   We find
that  scaling relations  such  as the  Kormendy relation,  photometric
plane and other correlations  involving bulge and disk parameters show
a luminosity dependence.  This dependence can be explained in terms of
galaxy  formation models in  which faint  lenticulars ($M_T  > -24.5$)
formed  via  secular  formation   processes  that  likely  formed  the
pseudobulges  of late-type disk  galaxies, while  brighter lenticulars
($M_T <  -24.5$) formed through  a different formation  mechanism most
likely involving  major mergers.  On probing variations  in lenticular
properties as  a function of  environment, we find that  faint cluster
lenticulars show  systematic differences  with respect to  faint field
lenticulars.  These differences support  the idea  that the  bulge and
disk  components fade  after the  galaxy falls  into a  cluster, while
simultaneously undergoing  a transformation from  spiral to lenticular
morphologies.

\end{abstract}


\begin{keywords}

galaxies: elliptical and lenticular -   fundamental parameters
galaxies: photometry - structure - bulges 
galaxies: formation - evolution
\end{keywords}


\section{Introduction}

Lenticular (S0) galaxies were  originally conceived as a morphological
transition class between ellipticals  and early-type spirals by Hubble
(1936). These galaxies have disks spanning a wide range in luminosity,
contributing between five to fifty  percent of the total galaxy light.
They are distinguished from spiral galaxies by the lack of conspicuous
spiral arms.   For these reasons, lenticulars  can be thought  of as a
population which  is intermediate  between ellipticals and  early type
spirals.  Indeed, the placement of lenticulars by Hubble (1936) on the
tuning fork  diagram clearly implies  such a relationship.  In several
observable  properties such  as bulge-to-disk  luminosity  ratio, star
formation rate  and colour,  lenticulars are, on  average, intermediate
between ellipticals and early type spirals.

\begin{table*}
\begin{center}
\begin{minipage}{166mm}
\caption{Best fit Bulge and Disk parameters for our sample.} 
\label{table1}
\begin{tabular}{l l l c c c c c c c c c}
\hline
Name & z & T  & \multicolumn{4}{c}{Bulge parameters} & \multicolumn{3}{c}{Disk parameters} & B/T & M$_T$ \\
\hline 
 & &  & $\mubzero$ & $\re$ & $n$ & $e_b$ & $\mudzero$ & $r_d$ & $e_d$ &  & \\  
 & &  & (mag arcsec$^{-2}$) & (kpc) &  &  & (mag arcsec$^{-2}$) & (kpc) &  &  &  \\ 
\hline
UGC\,00080 & 0.01019 &    -3 & 11.54 &  0.45 &  1.63 & 0.001 & 15.69 &  2.33 & 0.505 & 0.19 & -24.99  \\
UGC\,00491 & 0.01664 &    -1 & 10.38 &  4.95 &  3.82 & 0.257 & 17.65 &  5.08 & 0.231 & 0.65 & -25.53  \\
UGC\,00859 & 0.00711 &     0 & 11.34 &  1.39 &  3.36 & 0.141 & 16.40 &  1.22 & 0.403 & 0.43 & -23.25  \\
UGC\,00926 & 0.01467 &  -2.5 & 09.74 &  9.81 &  4.39 & 0.348 & 14.93 &  0.80 & 0.353 & 0.93 & -26.17  \\
UGC\,01250 & 0.01235 &    -2 & 09.51 &  1.92 &  3.79 & 0.397 & 15.92 &  2.66 & 0.696 & 0.32 & -25.24  \\
UGC\,01823 & 0.01332 &    -3 & 11.41 &  1.96 &  2.28 & 0.416 & 16.81 &  6.56 & 0.398 & 0.34 & -26.34  \\
UGC\,01964 & 0.01732 &    -2 & 10.06 &  4.03 &  3.96 & 0.103 & 15.60 &  1.86 & 0.585 & 0.58 & -25.33  \\
UGC\,02039 & 0.01505 &    -2 & 10.18 &  1.92 &  3.37 & 0.007 & 17.19 &  5.52 & 0.644 & 0.29 & -25.52  \\
UGC\,02187 & 0.01605 &    -2 & 13.11 &  0.45 &  0.76 & 0.276 & 15.13 &  1.99 & 0.554 & 0.14 & -25.17  \\
UGC\,02322 & 0.01447 &    -1 & 10.51 &  4.18 &  3.86 & 0.008 & 16.84 &  2.08 & 0.015 & 0.75 & -24.89  \\
UGC\,03178 & 0.01578 &    -2 & 09.84 &  2.55 &  3.89 & 0.014 & 14.76 &  1.17 & 0.479 & 0.47 & -24.90  \\
UGC\,03452 & 0.01876 &    -2 & 10.53 &  3.31 &  3.47 & 0.242 & 15.91 &  2.89 & 0.516 & 0.46 & -25.70  \\
UGC\,03536 & 0.01564 &    -2 & 13.18 &  0.55 &  0.99 & 0.012 & 14.80 &  2.03 & 0.576 & 0.10 & -25.48  \\
UGC\,03567 & 0.02016 &    -2 & 09.96 &  4.87 &  4.15 & 0.062 & 15.99 &  1.06 & 0.503 & 0.87 & -25.04  \\
UGC\,03642 & 0.01500 &    -2 & 11.94 &  1.03 &  2.02 & 0.010 & 17.74 &  8.45 & 0.206 & 0.17 & -25.72  \\
UGC\,03683 & 0.01908 &    -2 & 12.13 &  1.56 &  1.83 & 0.283 & 17.01 &  5.69 & 0.171 & 0.37 & -25.91  \\
UGC\,03699 & 0.01947 &    -2 & 10.35 &  3.13 &  3.32 & 0.336 & 16.54 &  1.65 & 0.489 & 0.87 & -25.36  \\
UGC\,03792 & 0.01908 &     0 & 10.17 &  4.41 &  3.74 & 0.224 & 18.15 &  5.71 & 0.225 & 0.72 & -25.66  \\
UGC\,03824 & 0.01787 &    -2 & 11.40 &  1.52 &  2.71 & 0.003 & 17.43 &  3.18 & 0.002 & 0.52 & -24.52  \\
UGC\,04347 & 0.01494 &    -2 & 10.68 &  4.33 &  3.34 & 0.243 & 16.34 &  1.01 & 0.528 & 0.95 & -25.57  \\
UGC\,04767 & 0.02413 &    -2 & 09.05 &  7.65 &  4.74 & 0.105 & 14.98 &  0.27 & 0.001 & 0.99 & -25.55  \\
UGC\,04901 & 0.02810 &    -2 & 13.30 &  3.06 &  1.78 & 0.305 & 18.08 & 14.28 & 0.043 & 0.27 & -26.70  \\
UGC\,05292 & 0.00511 &    -2 & 09.22 &  0.26 &  3.08 & 0.008 & 17.28 &  2.01 & 0.282 & 0.21 & -23.07  \\
UGC\,06013 & 0.02190 &    -2 & 11.29 &  5.10 &  3.68 & 0.325 & 15.41 &  1.00 & 0.001 & 0.78 & -24.89  \\
UGC\,06389 & 0.00663 &    -2 & 11.71 &  3.00 &  3.36 & 0.378 & 13.88 &  0.21 & 0.334 & 0.89 & -23.75  \\
UGC\,06899 & 0.02249 &    -2 & 09.79 &  5.11 &  4.06 & 0.060 & 16.71 &  0.53 & 0.001 & 0.99 & -25.35  \\
UGC\,07142 & 0.00328 &    -2 & 10.48 &  0.35 &  2.22 & 0.219 & 15.96 &  1.32 & 0.503 & 0.33 & -23.66  \\
UGC\,07473 & 0.00414 &    -2 & 11.76 &  0.24 &  1.20 & 0.239 & 14.55 &  1.20 & 0.725 & 0.12 & -24.57  \\
UGC\,07880 & 0.00388 &    -3 & 11.50 &  0.68 &  2.39 & 0.330 & 14.29 &  0.63 & 0.747 & 0.34 & -23.74  \\
UGC\,08675 & 0.00361 &    -2 & 09.18 &  0.25 &  3.35 & 0.020 & 17.14 &  1.48 & 0.011 & 0.19 & -22.53  \\
UGC\,09200 & 0.01081 &    -2 & 12.33 &  1.35 &  2.23 & 0.117 & 17.83 &  3.10 & 0.001 & 0.57 & -24.15  \\
UGC\,09592 & 0.01791 &    -2 & 08.69 &  1.62 &  3.62 & 0.315 & 16.30 &  2.33 & 0.123 & 0.65 & -25.30  \\
UGC\,11356 & 0.00820 &  -2.5 & 11.64 &  0.47 &  1.64 & 0.080 & 15.16 &  1.46 & 0.020 & 0.24 & -24.57  \\ 
UGC\,11972 & 0.01463 &  -2.5 & 10.29 &  8.62 &  4.08 & 0.441 & 14.24 &  0.45 & 0.002 & 0.95 & -26.00  \\
UGC\,12443 & 0.01463 &    -2 & 10.83 &  3.34 &  3.44 & 0.097 & 19.36 &  0.04 & 0.002 & 1.00 & -24.61  \\
UGC\,12655 & 0.01727 &    -2 & 10.34 &  1.14 &  2.87 & 0.191 & 16.86 &  4.77 & 0.400 & 0.24 & -25.47  \\

\hline
\end{tabular}

Notes.\ Column (1) gives the UGC catalogue  number, column (2) and  column (3)  
give redshift and morphological  type respectively from NED, columns (4), (5), 
(6) and (7) give unconvolved bulge central surface brightness, bulge effective 
radius, S{\'e}rsic index and bulge ellipticity respectively,  columns (8)  
(9) and (10)  give unconvolved disk  central   surface  brightness,   
disk  scale  length   and  disk ellipticity  respectively,   column  (11)  
gives   the  bulge-to-total luminosity ratio and  column (12) gives the 
absolute  magnitude in the $K$ band.
\end{minipage}
\end{center}
\end{table*}


Detailed study  of individual  lenticular galaxies indicates  that the
situation is more complex in  reality.  It has been suggested (van den
Bergh 1994) that there are different, but overlapping, sub-populations
amongst  the  lenticulars.  Exploration  of  formation  scenarios  for
lenticulars using  theory and numerical simulations  also suggest that
lenticulars  may have  formed in  different  ways.  They  could be  of
primordial  origin forming rather  rapidly at  early epochs,  or could
have  been formed by  the slow  stripping of  gas from  spirals, which
changes the  morphology (Abadi, Moore  \& Bower 1999), or  through the
mergers of unequal-mass spirals (Bekki 1998).  The two main components
of lenticulars  -- the bulge and the  disk -- may have  their own {\it
different}     and     possibly     {\it    independent}     formation
history. Observationally, the disks seem to be younger than the bulges
in both  spiral and lenticular  galaxies (Peletier \&  Balcells 1996),
with  the  age  difference   between  the  two  components  larger  in
lenticulars (Bothun \& Gregg 1990). Bars in lenticular galaxies are 
also of interest in many studies as they provide a clue about the evolutionary 
history of these galaxies. Recent studies reveal that bars in 
lenticular galaxies are  shorter, less massive and  have 
smaller bar torques (Buta \etal 2006; Laurikainen \etal 2005; 
Laurikainen \etal 2006; Gadotti, D. A. \etal 2007).

A detailed multiband study of the morphology  of  representative  samples  
of lenticulars  in  different environments,  and  comparison  of  their  
properties  with  those  of ellipticals,  and  with bulges  and  disks  
of  spirals will  be important  in  addressing   these  issues  observationally.   
However, comparing the predictions of models to observations is complicated 
for two  main reasons: (1)  The models  often use  simplifying assumptions
that may not  hold for real galaxies and (2) Models  often do not make
firm predictions about {\it directly observable} quantities. Nevertheless, 
the {\it statistical} properties of galaxy ensembles can be  compared  to 
model  predictions.  Such  a quantitative  comparison between  bulge and  
disk properties  of  galaxies is considerably simplified if  one assumes  
simple analytic profiles  for the  light distribution  of the
bulge  and  the  disk.  In  practice,  the  S{\'e}rsic  $r^{1/n}$  law
(S{\'e}rsic 1968)  adequately represents the  bulge surface brightness
distribution while  an  exponential  best  represents  the  disk
surface brightness distribution. The  total galaxy light is simply the
sum of a S{\'e}rsic bulge and an exponential disk.

In  this work,  we  obtain  and report  structural  parameters for  36
lenticular  galaxies observed  in the near-infrared  $K$ band.   Using 
results from the bulge-disk decomposition, we examine correlations among
the bulge  and disk parameters  and discuss implications to  models of
lenticular  formation. Our  goal is  mainly to  study  the constraints
placed on  bulge formation in  lenticular galaxies by  these parameter
correlations. Specifically, we want to investigate, in greater detail,
the  discovery   of  Barway  \etal  (2007)  who   found  two  distinct
populations of  bulges in lenticulars  that were fainter  and brighter
than a threshold luminosity.

This  paper   is  organised  as   follows:  the  sample,   details  of
near-infrared  observations,  data  reduction  technique and  photometric
calibration are   summarised  in   $\S$   2.   The two-dimensional decomposition 
analysis is described in $\S$ 3. In $\S$ 4, we discuss the  luminosity and 
environment dependence of lenticular galaxy properties.

Throughout this paper, we use the standard concordance cosmology 
with $\Omega_M = 0.3$, $\Omega_\Lambda = 0.7$, and $h_{100} = 0.7$.


\section{Observations and data reduction}
\subsection{Our sample}

Our sample  consists of a set  of 36 bright  field lenticular galaxies
from Barway \etal (2005). The original sample contained 40 lenticulars
galaxies,  selected from  the  Uppsala General  Catalogue (UGC),  with
apparent blue magnitude brighter than $m_B = 14$, diameter $D_{\rm 25}
< 3\arcmin$  and declination in  the range $5<\delta <  64^\circ$. The
sample,  while  not  complete,   is  representative  of  bright  field
lenticulars.

We obtained images of the sample galaxies in the near infrared $K$
band with the {\it Observatorio Astronomico Nacional} 2.1-m telescope
at San Pedro Martir, Mexico. The CAMILA instrument (Cruz-Gonzalez
\etal 1994), which hosts a NICMOS 3 detector of 256$\times$256 pixel
format, was used in the imaging mode. Each $K$ band observing sequence
consisted of 10 exposures, six on the object and four on the sky. The
net exposure times were, typically, 10 minutes per galaxy.  A series
of twilight and night-sky images were taken for flat-fielding
purposes.  The $K$ observations were carried out in four runs in
December 2000, March 2001, October 2001 and March 2002.  The data
reduction procedure for the $K$ images involved subtraction of the
bias and sky frames, division by flat field frames, registration of
the images to a common co-ordinate system and then stacking of all the
images of a given galaxy. All image reductions were carried out using
the Image Reduction and Analysis Facility (IRAF\footnote {IRAF is
distributed by National Optical Astronomy Observatories, which are
operated by the Association of Universities for Research in Astronomy,
Inc., under cooperative agreement with the National Science
Foundation.})  and the Space Telescope Science Data Analysis System
(STSDAS\footnote{STSDAS is a product of the Space Telescope Science
Institute, which is operated by AURA for NASA.}). Standard fields were
observed in order to enable accurate photometric calibration of our
$K$ band observations.  Reddening corrections due to Galactic
extinction and K-correction were applied to individual galaxies.  Full
details on sample selection, observation, and data reduction
procedures can be found in Barway \etal (2005).


\subsection{Comparison samples}

We supplement our  sample with data from Bedregal  \etal 2006 (hereafter
BAM06)    and    additional    data    provided   by    Bedregal    \&
Arag{\'o}n-Salamanca  in electronic  form (private  communication) who
used the  Two Micron All Sky  Survey (2MASS; \ Jarrett \etal 2003) data
for a structural analysis of a sample of 49 lenticular galaxies. These
are relatively  faint objects  with sufficient rotational  support for
the disks. Addition of these  data complements our sample in two ways:
(a) it provides a low luminosity  extension to the galaxies in our sample
and (b) it provides lenticulars  in different environments  \ie galaxies
from  the  Coma (14  galaxies),  Virgo  (8  galaxies), and  Fornax  (6
galaxies)  clusters  along  with  21 field  lenticulars.   Whenever  a
comparison with  other morphological  classes is appropriate,  we have
used results from analyses of the following samples: \\
\begin{itemize}
\item  A  sample of  42  elliptical  galaxies  from the  Coma  cluster
observed  in  $K$  band  by  Mobasher \etal  (1999)  as  analysed  by
Khosroshahi \etal (2000b).
\item A sample of 26 early  type spiral galaxies in the field observed
by  Peletier  \&  Balcells (1997)  in  the  $K$  band as  analysed  by
Khosroshahi \etal (2000b).
\item  A sample of 40  late-type spiral galaxies observed and analysed 
by M\"{o}llenhoff \& Heidt (2001) in the $K$ band. 
\end{itemize}

All the above analyses model the galaxy light using the S{\'e}rsic
function for the bulge and an exponential function for the disk. In
all cases, the bulge-disk decomposition is performed using the full 2D
image of the galaxy.  Incidentally, for all samples except that of
BAM06 and M\"{o}llenhoff \& Heidt (2001), the decomposition has been performed 
with the same code {\it fitgal} (Wadadekar \etal 1999). Details of the 
sample selection, observation, data reduction and bulge-disk decomposition 
of these samples may be found in the references cited.


\section{Analysis}
\subsection{2-d image decomposition}

Extracting  the  structural  parameters   of  a  galaxy  requires  the
separation  of the  observed light  distribution into  bulge  and disk
components.   There is considerable  variation in  the details  of the
decomposition techniques  proposed by various  researchers.  In recent
years, methods that  employ two dimensional fits to  broad band galaxy
images  have become  popular  (\eg \  Wadadekar, Robbason \&  Kembhavi
1999;  Peng \ \etal  2002; Simard  \etal 2002;  de Souza  \etal 2004).
Most  of  these   decomposition  techniques  assume  specific  surface
brightness distributions like the S{\'e}rsic  law for the bulge and an
exponential distribution for the disk.

Our decomposition procedure is a full two dimensional method that uses
information  from all pixels  in the  image. {\it  fitgal} essentially
involves a numerical solution  to a signal-to-noise ( hereafter $S/N$)
weighted $\chi^2$ minimisation  problem.  We achieve this minimisation
using the  Davidon-Fletcher-Powell variable metric  algorithm included
as  part of  MINUIT --  a multidimensional  minimisation  package from
CERN.  Our  technique involves  iteratively  building two  dimensional
image  models that  best  fit  the observed  galaxy  images, with  the
quality  of the  fit  quantified  by the  $\chi^2$  value. We  compute
weights for the $\chi^2$ function  using the $S/N$ ratio at each pixel
of the  galaxy image.  We convolve  the model image  with the measured
point spread  function from  the galaxy frame  before the  $\chi^2$ is
computed. Details of the accuracy and reliability of the decomposition
procedure as assessed by  simulations, are provided in Wadadekar \etal
(1999).

At  near-infrared wavelengths,  dust related  absorption  and emission
from patchy  star forming  regions are both  weak relative  to optical
wavelengths. The smooth, featureless light profiles of galaxies in the
near-infrared are  very convenient for modelling  using simple analytic
functions to represent the galaxy components.  Use of $K$ band data is
especially  important for  our sample,  as it  includes  several dusty
galaxies.   In our  bulge-disk  decomposition, we  have the  following
quantities  as  free   parameters:  (1)~$I_b(0)$:  the  central  bulge
intensity,  in counts,  which can  later  be converted  to $\rm  mag\,
arcsec^{-2}$  using the  photometric calibration  (2)~$r_e$:  the half
light radius of the bulge  in pixels (3)~$e_b$: the ellipticity of the
bulge (4)~$n$:  the bulge  S{\'e}rsic index (5)~$I_d(0)$:  the central
intensity of  the disk  in counts (6)~$r_d$:  the scale length  of the
disk in pixels (7)~$e_d$: the ellipticity of the disk

With these definitions, the S{\'e}rsic bulge intensity distribution can
be written as
\begin{eqnarray}
I_{bulge}(x,y)  &=& I_b(0) e^{  -2.303 b_n  (r_{bulge}/r_e)^{1/n}}, \\
r_{bulge} &=& \sqrt{x^2 + y^2/(1 - e_b)^2}, \nonumber
\end{eqnarray}  
where $x$  and $y$  are the  distances from the  centre of  the galaxy
along the major and minor axis respectively and $b_n$ is a function of
$n$  and  the root  of  an  equation  involving the  incomplete  gamma
function.  However, following Khosroshahi  \etal (2000b), $b_n$ can be
approximated as a linear function  of $n$, accurate to better than one
part in $10^{5}$, by
\begin{eqnarray}
b_n  &=& 0.868242\, n - 0.142058. \nonumber
\end{eqnarray} 
For $n=4$, which corresponds to de Vaucouleurs law, $b_4 = 3.33$.

The projected disk profile is represented by an exponential distribution,
\begin{eqnarray}
I_{disk}(x,y) &=& I_d(0) e^{- r_{disk}/r_d},\\
r_{disk} &=& \sqrt{x^2 + y^2/(1 - e_d)^2}. \nonumber
\end{eqnarray}
The ellipticity of the disk in the image is due to projection effects alone 
and is given  by
\begin{eqnarray}
e_d &=& 1 - \cos(i),
\end{eqnarray}
where $i$ is the angle of inclination between the line of sight and
the normal to the disk plane.

The bulge-to-disk luminosity ratio  is a dimensionless parameter which
is  commonly   used  as  a  quantitative   measure  for  morphological
classification of galaxies. For  a S{\'e}rsic bulge and an exponential
disk it is given by
\begin{equation}
(B/D)_n=\frac{n\Gamma                      (2n)}{(2.303b_n)^{2n}}\left(
\frac{I_b(0)}{I_d(0)}\right)      \left(     \frac{r_e}{r_d}\right)^2.
\labequn{dbybn}
\end{equation} 
We  use   the  bulge-to-total  luminosity  ratio   $B/T  =
(B/D)/(1+(B/D))$ in this paper, since it spans a restricted range of 0
to 1.


\begin{figure}
\centerline{\psfig{figure=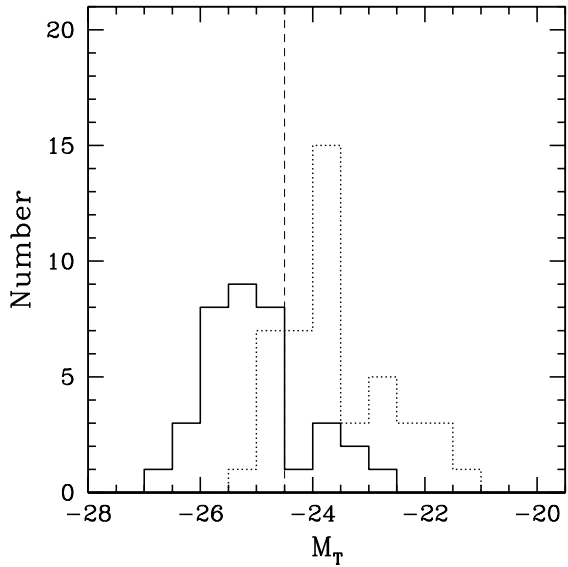,width=8.0cm,angle=0.}}
\caption{Distribution of  total absolute magnitude (M$_T$)  in $K$ band
for  our  sample  (solid  line)  and  for  BAM06  lenticulars  (dotted
line). The vertical  dashed line  corresponds to total  absolute magnitude
M$_T$  = -24.5,  which we  use  to divide  low- and  high-luminosity
lenticulars.}
\label{f1}
\end{figure}


\subsection{Extraction of structural parameters using 2-d image decomposition}

Although the $K$ band observations are advantageous for their relative
lack of  absorption related problems,  the high sky background  in $K$
band  poses  some  difficulty  in extracting  the  global  photometric
parameters accurately,  as it  limits the region  where $S/N$  is high
enough for  the pixels to  be usable.  Even  a small error in  the sky
estimation can lead to spurious  results because of the relatively low
$S/N$ in the  images. For example, if the  sky is under-estimated, say
by 5\%, the residual background can  lead to a spurious detection of a
large circular  disk with  a non-zero $\idzero$.   Even when  a `bulge
only' model is used, a wrong  assessment of the background can lead to
serious mis-estimation of the morphological parameters.

In our $K$ band images, the sky background is automatically subtracted
from the many individual sky frames for each galaxy during the
preprocessing stage (see Barway \etal 2005 for details).  However, if
sky is removed, noise statistics calculations at low $S/N$ (where the
sky dominates) are grossly inaccurate. To avoid this problem, we added
back the sky background that had been subtracted from each galaxy
image.

Our bulge-disk decomposition code \emph{fitgal} (Wadadekar \etal 1999)
allows  the  estimation  of  the background  simultaneously  with  the
estimation of  morphological parameters.  However,  such a measurement
can  lead to  unstable values  for the  parameters, as  the background
counts are much  higher than the signal even in  the central region of
the galaxies. To avoid this instability,  we had a run for each galaxy
in which the aim was to  estimate the background rather than to obtain
the morphological parameters accurately. To this end, we first ran the
code on a  selected region of each galaxy image,  dominated by the sky
background. We  compared the value  of the background returned  by the
fitting  procedure, with  that  estimated from  regions  of the  image
frames free of  the galaxy and from the observed  sky images to ensure
that these values are all consistent  with each other. We then adopted the
sky value returned  by the code as the best  representation of the sky
background, since  this uses information from all  the relevant pixels
of the image, and fixed  the background parameter to this value during
all subsequent runs of \emph {fitgal}.

We fitted ellipses to the isophotes  of each galaxy of our sample, and
obtained  the one  dimensional  surface brightness  profile along  the
major axis, using  the STSDAS task \emph{ellipse}. For  this task, sky
subtracted galaxy  frames were used.  The fits provided us  with model
independent surface brightness  distributions, which could be compared
to the surface brightness distribution predicted by our best fit bulge
plus disk models.  In order to obtain good initial  values for our two
dimensional fits, we fitted the inner part of one dimensional profiles
obtained above with de Vaucouleurs  law profiles (taking care to avoid
the  region  affected  by  the  PSF)  and  the  outer  part  with  the
exponential law. This provided us with approximate values of $\ibzero$
and $\re$ for bulge and $\idzero$ and $\rd$ for disk, which we used as
input to obtain a full two dimensional S{\'e}rsic and exponential model fit
to each galaxy in the sample. During the fit for each galaxy, we fixed
the background  to the  value obtained as  described above. Of  the 40
galaxies in our  sample, we could not get  satisfactory fits for three
galaxies,  UGC\,03087,  UGC\,11178  and  UGC\,11781. UGC\,03087  is  a
strong radio source with an optical  jet visible in the image make the
fit  unreliable. For  galaxies UGC\,11178  and UGC\,11781,  our images
suffer from poor  $S/N$ ratio and we are not  able to get satisfactory
fits.  UGC\,07933   is  classified  as   an  elliptical  in   the  RC3
catalogue. We  therefore exclude  these four galaxies  from subsequent
discussions, leaving us with a  sample of 36 galaxies.  We have listed
in  Table 1 the  best fit  bulge and  disk parameters  for all  the 36
lenticular galaxies in the sample.

BAM06 used images in the $K$ band from the 2MASS survey to obtain
bulge and disk parameters for the 49 galaxies in their sample, using
the {\it gim2d} decomposition code by Simard \etal (2002), assuming
S{\'e}rsic and exponential laws for the bulge and disk light
distributions, as we have done with our sample.  Barway \etal (2007)
have shown (see their Figure 2a) that there are three clear outliers
among the cluster lenticulars of the BAM06 sample.  Inspection of the
2MASS $K$ band images of the corresponding galaxies, shows that one of
these (ESO\,358-G59) has poor $S/N$ ratio, while the other two
(NGC\,4638 and NGC\,4787) are obviously disk-dominated systems, which
are likely to have disk scale lengths larger than those reported by
BAM06.  Following Barway \etal (2007), we omit these three outliers
from further discussion, while noting that our conclusions are not
significantly changed by this omission.

It must be noted that the depth of our $K$ band images is considerably
more than the depth of 2MASS images used by BAM06. We have a typical
exposure time of 10 minutes per galaxy as opposed to 7.8 sec/frame in
2MASS.  This combined with the fact that we used a somewhat
larger telescope for obtaining our data allows us to reach a depth of
typically 22 mag arcsec$^{-2}$, corresponding to an error in
photometry of 0.1 mag arcsec$^{-2}$. The 2MASS data used by BAM06
reach a typical depth of 20 mag arcsec$^{-2}$ for the same error. This
implies that our images are about 2 magnitudes deeper than the 2MASS
data. Given the significant difference in typical depth, bulge disk
decompositions of the low S/N 2MASS data may, in principle, be
systematically affected. To probe the extent of this effect, we
obtained 2MASS images for all the galaxies in our sample and performed
the decomposition using our technique on these data. For 80\% of
galaxies, we found a reasonable match between parameters extracted
from our data and the 2MASS data. For 20\% of galaxies our fitting
procedure failed when using 2MASS data, indicating that the low S/N
was affecting our ability to extract parameters reliably for these
galaxies. However, such an inability to obtain reliable fits is
unlikely to affect the parameter extraction procedure of BAM06, which
is based on the Metropolis algorithm (Metropolis \etal 1953), as
implemented in the {\it gim2d} software (Simard \etal 2002). {\it
gim2d} is a robust, albeit slow, bulge-disk decomposition software
that obtains the global minimum of a fit, in almost any situation. It
is thus well suited for parameter extraction for relatively shallow
data like that of BAM06, where the low S/N is likely to make fitting
difficult.
 
Another possible concern is whether a linear combination of a bulge
and a disk is an adequate mathematical formulation to model the light
distribution of galaxies in our sample e.g. if a significant
non-axisymmetric bar (or similar structure) were present for the
galaxies in our sample, that would affect the parameters that we
extract (see Laurikainen et al. 2005 for a discussion on the fitting
of non-axisymmetric components). It is fortunately easy to detect the
presence of non-axisymmetric structure by examining the residuals of
our fitting, which are computed as the difference between the galaxy
image and best fit model. Since the analytic functions we use to model
the bulge and disk are axisymmetric, the bars are not modeled in our
scheme. They remain in the residual image, and can be visually
seen if they are bright enough. For the 36 galaxies in our sample, the RC3 
catalog indicates that 5 may have bars. We find from the residuals that 
only one galaxy (UGC 80) has a discernable bar. For the BAM06 galaxies, 
the RC3 catalog indicates that 14 out of 47 galaxies may have bars. We 
find non-axisymmetric structures in 11 of these 15 galaxies. For these 
galaxies, extracted parameters are likely to be somewhat inaccurate. For 
the analysis subsequently reported in this paper, such inaccuracy is of 
concern only if it leads to {\it systematic} offsets between barred 
and non-barred galaxies. We demonstrate in Section 4.1.1 that such obvious 
systematic offsets do not exist, implying that inaccurate parameter estimation 
for a small fraction of galaxies will not alter the results of this work. 


\begin{figure}
\centerline{\psfig{figure=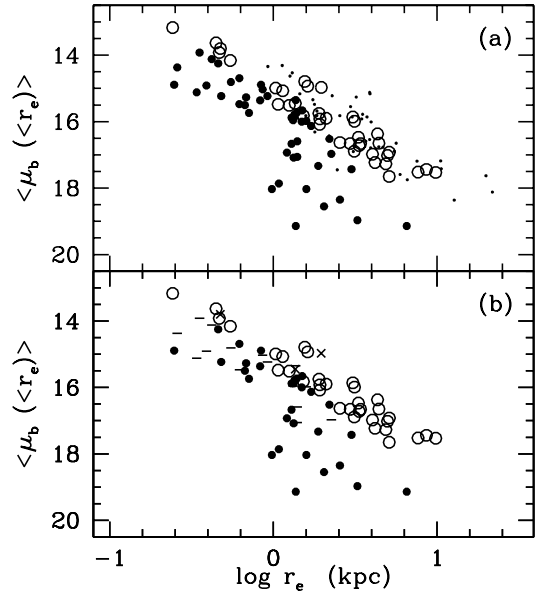,width=8.0cm,angle=0.}}
\caption{Dependence  of  the   mean  surface  brightness  within  bulge
effective radius ($\mmubre$) on bulge effective radius $\re$  (a) for
bright (as  open circles) and  faint lenticulars (filled circles).
Coma ellipticals (as dots) are overplotted for comparison. (b) for bright 
lenticulars with bars (as crosses) and faint lenticulars with bars (as dashed) 
as classified in RC3 catalogue.}
\label{f2}
\end{figure}

\begin{figure}
\centerline{\psfig{figure=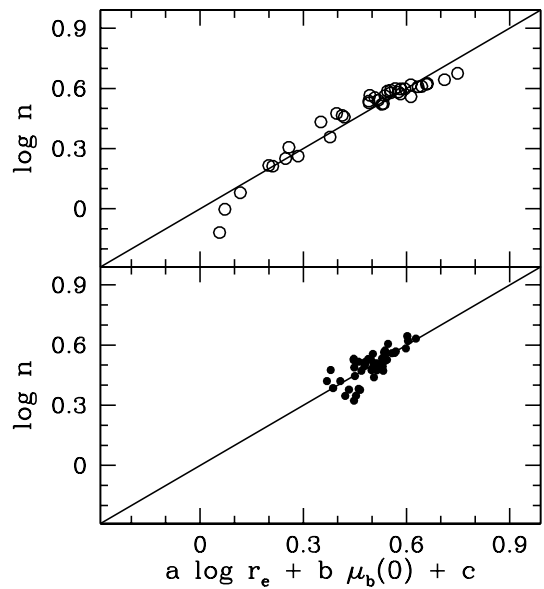,width=8.0cm,angle=0.}}
\caption{An   edge-on   view    of the photometric plane   (PP)   for
lenticulars.   Top:  PP  for   bright  lenticulars;   {\rm  log n  =
0.251$\pm$0.045   log $\re$   -   0.094$\pm$0.012    $\mubzero$   +
1.377$\pm$0.138}.  Bottom: PP  for  faint lenticulars;  {\rm  log n  =
0.185$\pm$0.029   log $\re$   -   0.041$\pm$0.004    $\mubzero$   +
0.935$\pm$0.046}.}
\label{f3}
\end{figure}

\begin{table*}
\begin{center}
\begin{minipage}{120mm}
\caption{photometric plane  coefficients  for bright  \&  faint lenticulars. 
}
\label{table2}
\begin{tabular}{c c c c c c c}
\hline
Lenticular Sample &  a & b & c &  N & $\,{\rm  rms}_{n}$ & $\,{\rm rms}_{o}$ \\  
\hline
Bright & 0.251$\pm$0.045 & $-$0.094$\pm$0.012 &  1.377$\pm$0.138 & 38 & 0.037 & 0.036 \\   
Faint  & 0.185$\pm$0.029 & $-$0.041$\pm$0.004 &  0.935$\pm$0.046 & 44 & 0.039 & 0.038 \\  
Bright with restricted  n & 0.150$\pm$0.0213 & $-$0.056$\pm$0.007 & 1.050$\pm$0.067 & 29 & 0.019 & 0.019 \\
\hline
\end{tabular}

Notes. \ a and b are the coefficients of $\log r_{\rm e}$ and $\mu _{\rm b}(0)$ 
respectively, c is the constant, N is the number of galaxies, $\,{\rm  rms}_{n}$ is 
the r.m.s. scatter measured along  the $\log  n$ axis  and $\,{\rm  rms}_{o}$ is  
the  r.m.s. scatter measured in a direction orthogonal to the best-fit photometric plane.
\end{minipage}
\end{center}
\end{table*}


\section{Correlations and discussion}
\subsection{Luminosity dependence}

Barway \etal \ (2007) have presented evidence to support the view that
the formation  history  of  lenticular  galaxies depends  upon  their
luminosity.  According to  this view,  low-luminosity lenticular galaxies likely
formed  by the  stripping of  gas from  the disk  of  late-type spiral
galaxies, which in turn  formed their bulges through secular evolution
processes. On the other  hand, more luminous lenticulars likely formed
at early epochs through a  rapid collapse followed  by rapid star formation.

As we mentioned  earlier, the BAM06 galaxies complement  our sample in
two ways: they extend  to fainter luminosities and provide lenticulars
in  different  environments.    In  Figure~\ref{f1}  we  show  the
distribution of total  absolute magnitude ($M_T$) in the  $K$ band for
the combined sample, which is seen to span a wide range in luminosity.
We divide the combined sample into faint and bright groups, using $M_T
= -24.5$  as a  boundary.  The bright sample has 37 lenticulars while 46 
lenticulars belong to the faint sample according to this luminosity division.
The  boundary at $M_T  = -24.5$  is somewhat arbitrary but our results do  
not critically depend on small shifts in the dividing luminosity.  For 
instance,  changing this value by half a magnitude on either side  maintains  
correlations significant at least at the 95\% level. 

Using the luminosity division at $M_T  = -24.5$, Barway \etal \ (2007) reported  markedly 
different correlations between bulge effective radius ($\re$) and disk  
scale length ($\rd$) for bright and faint lenticulars. A  positive correlation 
of bulge and  disk sizes is
expected  if  the bulge  grows  over  time  through secular  evolution
processes  (see  review  by  Kormendy  \&  Kennicutt  2004).  No  such
correlation  is  expected  if  the  bulge formed  via  merger  related
processes.  Barway \etal \ (2007) found this positive correlation between
bulge and disk sizes of low luminosity lenticulars; such a correlation
was not observed for bright lenticulars. Moreover, they found that the
correlation holds for faint galaxies, irrespective of whether they are
situated in a field or cluster environment.

If  the differences between  low and  high luminosity  lenticulars are
indeed fundamental,  then there should be  systematic differences seen
between  the two  populations  in other  correlations among bulge and  
disk parameters. We study such correlations in the following sections.


\begin{figure}
\centerline{\psfig{figure=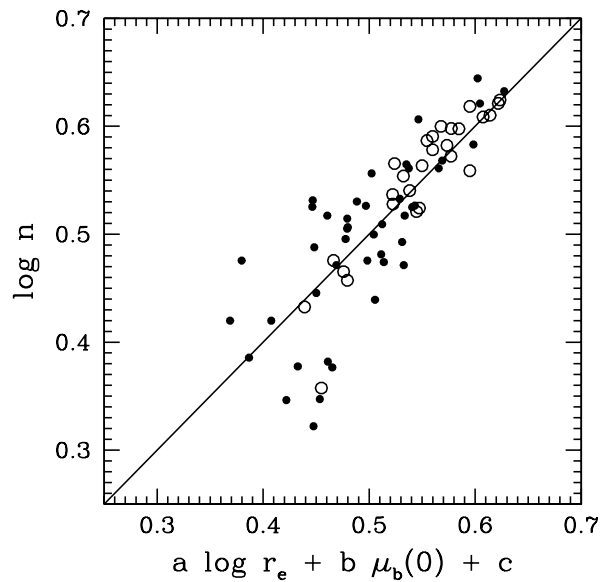,width=8.0cm,angle=0.}}
\caption{Photometric Plane for bright lenticulars (open circles) with
restricted  $n$ value. Faint  lenticulars (filled  circles) are over plotted 
on the same photometric plane. These show a higher scatter than the bright ones.}
\label{f4}
\end{figure}


\subsubsection{Kormendy relation and the effect of bars}

For samples of large elliptical galaxies, the central surface
brightness $\mubzero$ (or, equivalently, the mean surface brightness
($\mmubre$) within $\re$) is correlated with $\log \ \re$.  This was
first reported by Kormendy \ (1977) and is known as the Kormendy
relation.  In Figure~\ref{f2}(a) we show the Kormendy relation for
bright (as open circles) and faint lenticulars (as filled
circles). The bright lenticulars show tight correlation with Pearson
correlation coefficient $r = 0.95$ with significance greater than
99.99 \%.  The Kormendy relation for faint lenticulars shows
considerably greater scatter although the Pearson correlation
coefficient is $0.75$ with significance greater than 99.99
\%. Figure~\ref{f2}(b) show the Kormendy relation for bright (as
crosses) and faint (as dashes) lenticulars classified as barred
lenticulars in RC3 catalogue to probe for any systematic effect caused
due to presence of bars. It is clear from the figure that barred and
non-barred show no systematic offset. Only three galaxies are
classified as barred in our bright galaxy sample (indicated by
crosses). Even after removing these three galaxies, the correlation is
nearly unchanged with Pearson correlation coefficient $r = 0.96$ with
significance greater than 99.99 \%.  In the faint lenticular sample,
there are as many as 16 galaxies classified as barred. Nevertheless,
neglecting these 16 galaxies does not affect the correlation which has
$r = 0.74$ with significance greater than 99.99 \%. For all other
correlations subsequently reported in this paper, we find that
excluding barred galaxies does not significantly alter the correlation
coefficient or its significance. Hence, in further discussion, we do
not differentiate between barred and non-barred lenticulars, and
include both types in the analysis.

The Kormendy relation is known  to be a projection  of the fundamental  
plane (Djorgovski \& Davis 1987; Dressler \etal 1987) of galaxies.  A 
tight  Kormendy relation indicates  
(near) virialized bulges like  those found  in elliptical galaxies.   In 
Figure 2a, ellipticals  from  the  Coma  cluster  are overplotted  
as  dots.   As expected,  these ellipticals  show  a tight  correlation with  
Pearson correlation  coefficient $r  =  0.81$ with  significance greater  than
99.99  \%. The  slopes of  the best-fit  Kormendy relation  for bright
lenticulars  and Coma  ellipticals are similar.   This similarity coupled 
with the systematic difference  in scatter  in the  Kormendy relation  between  
bright and faint lenticulars is an indicator  of the more virialized state 
of the bright lenticular bulges and their close relation to ellipticals.


\subsubsection{Photometric Plane}

Analogous  to the  fundamental plane (FP),  ellipticals and  the  bulges of
early type spiral  galaxies obey a single planar  relation of the form
$\log \ n  = a \ \log \ r_{\rm e}+ b \ \mu_{\rm b}(0)+ c $  which Khosroshahi
\etal  (2000a,b)  called the  photometric  plane  (PP). Assuming  that
galaxy bulges behave  as spherical, isotropic, one-component systems,
there is a  unique specific entropy that can  be associated with every
galaxy  (Lima-Neto \etal  1999).   This in  turn,  implies a  relation
between  bulge   parameters  that   is  equivalent  to   the  observed
photometric plane. An edge-on view of the best-fit PP is shown in the
top  and bottom  panels of  Figure~\ref{f3} for  bright  and faint
lenticulars respectively. Table 2  lists the coefficients for best-fit
PP relation.  It is  immediately evident from Figure~\ref{f3} that
the PP for faint lenticulars spans a limited range of $n$. We restrict
our  bright  lenticular sample  to  the  range  spanned by  the  faint
lenticulars  and then  obtain  the  best fit  PP  for this  restricted
sample.   The rms  scatter  for the  (restricted)  bright galaxies  is
considerably smaller than  that for the faint galaxies  (see Table 2).
In  Figure~\ref{f4}  we  show  the   PP  for  these  two  sets  of
galaxies. Visually, the higher scatter of the faint galaxies about the
PP is obvious. The PP is a natural state for galaxies that formed like
ellipticals (at early epochs in  a burst of rapid star formation, with
stellar orbits that are well  relaxed). The fact that bulges of bright
lenticulars  have  a  tight  PP  is  an  indicator  of  homology  with
ellipticals. Faint lenticulars seem to belong to a separate class.


\begin{figure}
\centerline{\psfig{figure=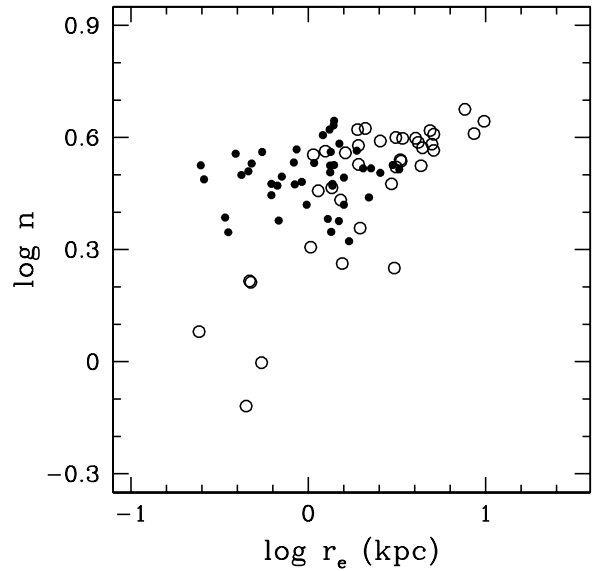,width=8.0cm,angle=0.}}
\caption{S{\'e}rsic index $n$ plotted against bulge effective radius
$r_e$ in kpc.  Symbols are as in Figure 2.}
\label{f5}
\end{figure}


\subsubsection{Correlations involving the S{\'e}rsic index ($n$)}

\begin{figure}
\centerline{\psfig{figure=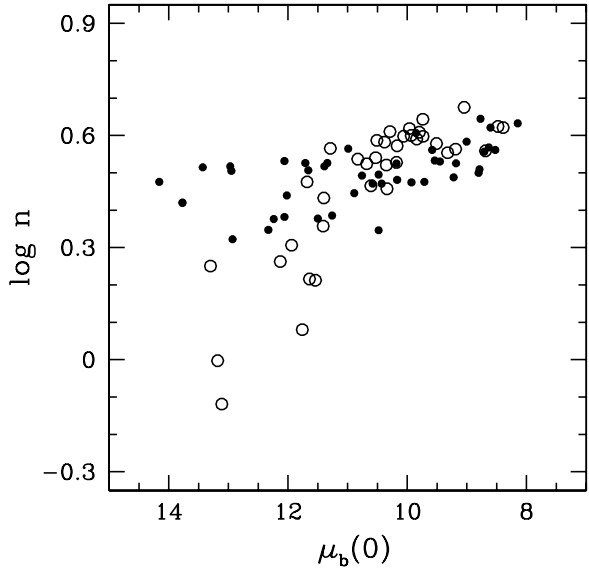,width=8.0cm,angle=0.}}
\caption{S{\'e}rsic  Index $n$  as  a function  of unconvolved  bulge
central surface brightness. Symbols are as in Figure 2.}
\label{f6}
\end{figure}

The  restricted  range  spanned  by  the  S{\'e}rsic  index  in  faint
lenticular  galaxies also  affects other  correlations  involving this
index.  Figure~\ref{f5} is a plot of the S{\'e}rsic index ($n$) as
a function  of effective radius  ($\re$) which again  shows systematic
differences  between  bright  and   faint  lenticulars.   Bright
lenticulars   are  well  correlated  having   Pearson  correlation
coefficient $r  = 0.79$  with significance greater  than 99.99  \% but
faint  lenticulars do not  show a  significant correlation.   A strong
correlation  exists   between  $n$  and  the   bulge  central  surface
brightness  ($\mubzero$),  as  shown  in Figure~\ref{f6}  for  the
brighter lenticulars. The Pearson correlation coefficient in this case
is 0.84 at a significance level better than 99.99\%. Faint lenticulars
also exhibit  good correlation having  Pearson correlation coefficient
$r =  0.58$ at  significance 99.96 \%,  but have a  markedly different
slope.  Comparing Figure~\ref{f5} and Figure~\ref{f6} 
with Figure 3 and Figure  4 respectively of Khosroshahi \etal (2000b),
we observe a  close correspondence in the distribution  of the bulges of
{\it bright} lenticulars and early  type spirals. Bulges of early type
spirals,  in  turn,  have  a  close  correspondence  with  ellipticals
(Khosroshahi \etal  2000b). On the other hand,  faint lenticulars seem
to be a different population.

\begin{figure}
\centerline{\psfig{figure=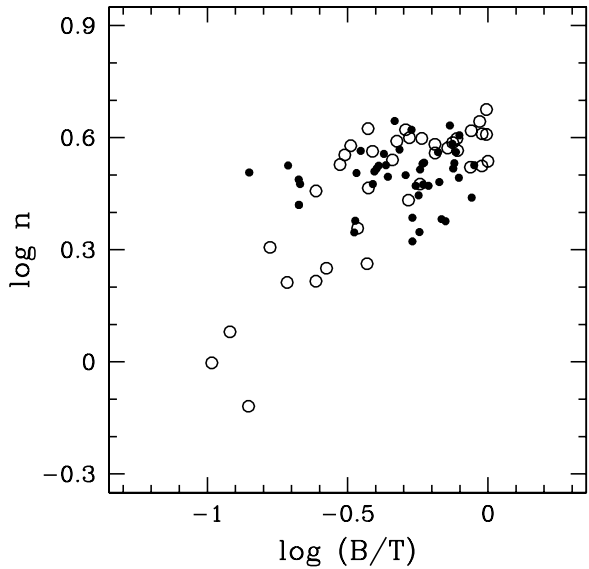,width=8.0cm,angle=0.}}
\caption{S{\'e}rsic   Index   $n$   vs.   bulge-to-total   luminosity
ratio. Symbols are as in Figure 2.}
\label{f7}
\end{figure}

The  bulge-to-total  luminosity  ratio  ($B/T$)  increases  with
S{\'e}rsic  index $n$  as  shown in  Figure~\ref{f7} for  brighter
lenticulars  ($r   =  0.81$  with  significance   greater  than  99.99
\%). Faint lenticulars show a weak correlation between $ \log \ (B/T)$ and
$\log \ n$.  Again, the  trends for  bright lenticulars  are consistent
with those  found by Khosroshahi \etal (2000b) in early-type spirals.  Bright 
lenticulars plotted in Figure 3 of Barway \etal (2007) include 5 galaxies that are 
obvious outliers in the anti-correlation of bulge and disk sizes. These bright 
lenticulars viz. UGC\,80, UGC\,2187, UGC\,3536, UGC\,7473 and UGC\,11356  do follow 
the correlations involving the S{\'e}rsic index. However, they are all characterised 
by a low S{\'e}rsic index, low bulge central surface brightness,  low bulge effective 
radius and low bulge-to-total luminosity ratio. This places these lenticulars in the 
bottom left quadrant of Figures 5, 6 and 7. Their bulge/disk parameters indicate that 
these are disk-like systems, with relatively weak bulges, making them unlikely 
candidates of major merger induced formation. Nevertheless, they do follow the same 
correlations, involving the S{\'e}rsic index, as the other bright galaxies. This 
discrepancy needs to be explored with a larger sample of bright, but disky lenticulars.


\begin{figure}
\centerline{\psfig{figure=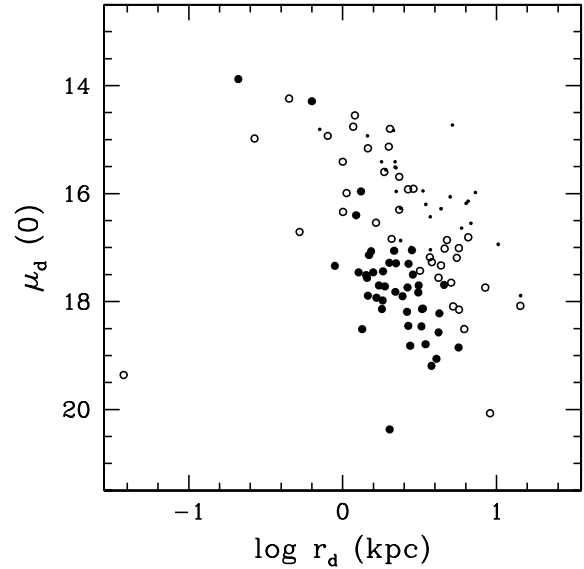,width=8.0cm,angle=0.}}
\caption{Unconvolved disk central  surface brightness $\mu_d(0)$ as a
function of  disk scale length. Early-type  spirals (as dots) are
overplotted for comparison. Other symbols are as in Figure 2.}
\label{f8}
\end{figure}


\begin{figure}
\centerline{\psfig{figure=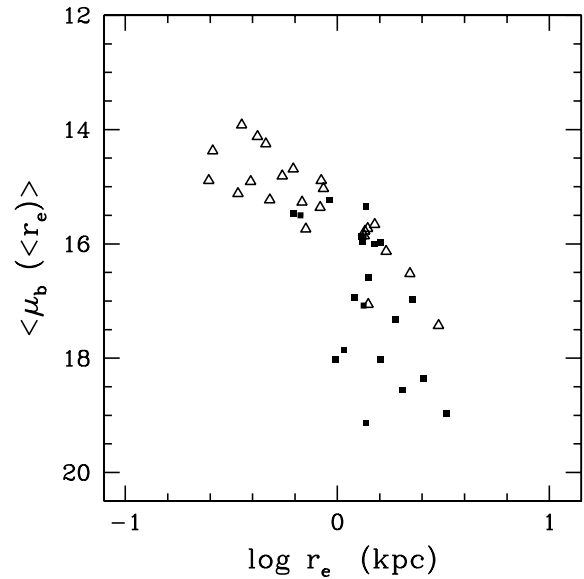,width=8.0cm,angle=0.}}
\caption{Dependence  of  the   mean  surface  brightness  within  bulge
effective radius ($\mmubre$) on  bulge effective radius $\re$ as
a  function of  environment.  Faint  field lenticulars are shown  as open
triangles   while  faint   cluster  lenticulars are  plotted   as  filled
squares.}
\label{f9}
\end{figure}


\subsubsection{Disk correlations}

There exist  clear correlations between  the two main  disk parameters
--central  surface  brightness  of  the disks  ($\mudzero$)  and  disk
scale-length  ($\rd$)-- \ for  both  bright and  faint lenticulars  (see
Figure~\ref{f8})  although they  occupy different  regions  of the
plot.   Lenticulars with  large  disks have  a  lower central  surface
brightness,  on average.   A  clear anti-correlation  is seen  between
these two  disk parameters for  bright lenticulars but the  scatter is
large.  The linear correlation  coefficient between  $\mudzero$ \ and log
$\rd$ is 0.43 with a  significance of 99.37 percent. Faint lenticulars
show  less scatter  with  the  correlation coefficient  of  0.70 at  a
significance  level  better  than  99.99  \%.   Similar  statistically
significant correlation was reported  by Khosroshahi \etal (2000b) for
early-type  spirals  and  M\"{o}llenhoff \& Heidt  (2001)  for  late-type
spirals for  these disk  parameters. In Figure~\ref{f8},  we also
plot as dots the central surface  brightness of the disks ($\mudzero$) against
the disk scale length ($\rd$)  of galaxies from the $K$ band observations of
Khosroshahi \etal  (2000b) of early-type  spirals.   From the
plot it is apparent that the  disk scale lengths of our sample and the
Khosroshahi \etal  (2000b) sample span the same  range, but $\mudzero$
are  brighter,  on  the  average,  for  early-type  spirals  than  faint
lenticulars.


\begin{figure}
\centerline{\psfig{figure=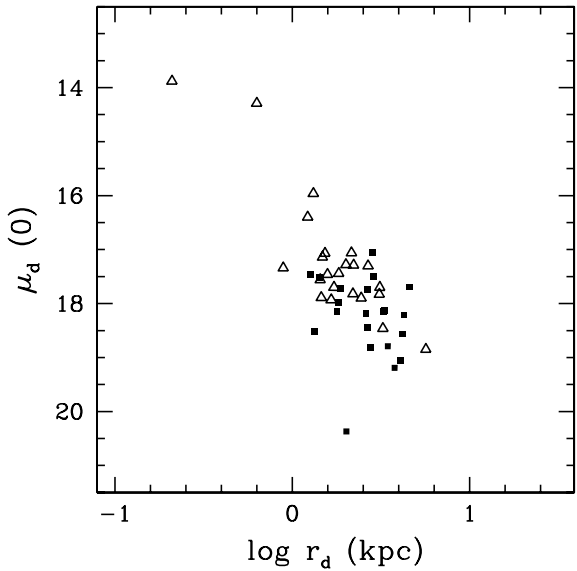,,width=8.0cm,angle=0.}}
\caption{Unconvolved disk  central surface brightness  $\mudzero$ is
plotted   against  disk   scale  length   $\rd$  as   a   function  of
environment.  Symbols are as in Figure 9.}
\label{f10}
\end{figure}


\subsection{Environmental dependence}

The dependence  of cluster environment  on formation and  evolution of
galaxies  is well  known. The  environment  influence is  best visible  through the
well-known   morphology-density  relationship,   according   to  which
early-type  galaxies  are preferentially  found  in rich  environments
(Dressler 1980),  with spirals are found predominately  in low density
environments.    One   of   the    stronger   trends   seen   in   the
morphology-density relation  at low redshift is  the dramatic increase
in the number of lenticular galaxies in rich clusters (Dressler \etal 1997),
suggesting    that   the    dominant   process    in    defining   the
morphology-density relation  is the transformation  of spiral galaxies
into lenticular  galaxies within rich clusters  (Poggianti \etal 1999;
Kodama \& Smail 2001). These spiral galaxies are continuously supplied
by  accretion from  the surrounding  field  during the  course of  the
assembly of  the cluster.  {\it Hubble Space  Telescope (HST)} imaging
of  10 clusters  at higher  redshift (z  $\sim$\,0.3-0.5)  by Dressler
\etal  1997  shows  similar  behaviour.  The  mechanisms  proposed  for
morphological  transformation   of  spirals  to   lenticulars  include
dynamical interactions  such as  galaxy harassment (Moore  \etal 1996,
1998, 1999)  and gas dynamical processes e.g.   ram pressure stripping
(Abadi \etal 1999). Simulations have shown that galaxy harassment is a
very efficient  mechanism for transforming early-type  disk galaxies to
lenticulars (Moore 1999) with consequent reduction in disk sizes. Ram 
pressure  stripping truncates the  star formation by  removing the
cold  neutral gas  reservoir causing the  disk component  of a spiral 
galaxy to fade within the cluster environment.

To  address   these possibilities   observationally,  it  is  necessary  to
investigate how scaling relations defined by various galaxy properties
vary  between field  and cluster  environments.  In  this  section, we
examine     a     few     correlations     as    a     function     of
environment.   Unfortunately,  we   are  unable   to   include  bright
lenticulars in this comparison because  only 3 cluster galaxies out of
the 38  in our sample,  are bright. For  faint galaxies, the  number of
galaxies  in the  field and  in clusters  are almost  equal,  making a
comparison possible.

The   Kormendy    relation   for   faint    lenticulars   plotted   in
Figure~\ref{f2}   shows  two   distinct   groups,  unlike   bright
lenticulars. To investigate the origin  of this dichotomy in the faint
lenticulars,  we plot  the  Kormendy relation  for faint  lenticulars,
differentiating  them  by   environment,  in  Figure~\ref{f9}.  We
indicate field lenticulars with open triangles and cluster lenticulars
with  filled  squares.   The  field  lenticulars seem  to  follow  the
Kormendy relation reasonably well while the cluster lenticulars show a
clear downward scatter with respect  to the relation. On average, this
requires a 1.5 mag arcsec$^{-2}$ fading of the mean surface brightness
within  $\re$  of  faint  cluster  lenticulars, with  respect  to  the
Kormendy  relation.  Boselli  \&  Gavazzi (2006)  have suggested  that
low-luminosity  lenticulars in  clusters might  be the  result  of ram
pressure stripping of late-type galaxies, which causes fading and thus
lowers surface brightness.

It is interesting to see if this fading is also seen for the disk
component.  In Figure~\ref{f10}, we plot the disk central surface
brightness against the disk scale length. As above, we indicate faint
field and cluster lenticulars with different symbols.  The faint field
lenticulars exhibit a clear anti-correlation, whereas the cluster
lenticulars occupy a limited region of the plot and show a downward
scatter (indicating fading of the disk), very similar to the scatter
seen in Figure~\ref{f9}.  Our results obtained using only photometric
data are consistent with the spectroscopic results of Barr \etal
(2007), which support the theory that lenticular galaxies are formed
when gas in normal spirals is removed (from both bulge and disk),
possibly when well formed spirals fall into the cluster.  It must be
noted that the sample of Barr \etal (2007) is a subset of the BAM06
sample, which is included in our analysis.

\begin{figure}
\centerline{\psfig{figure=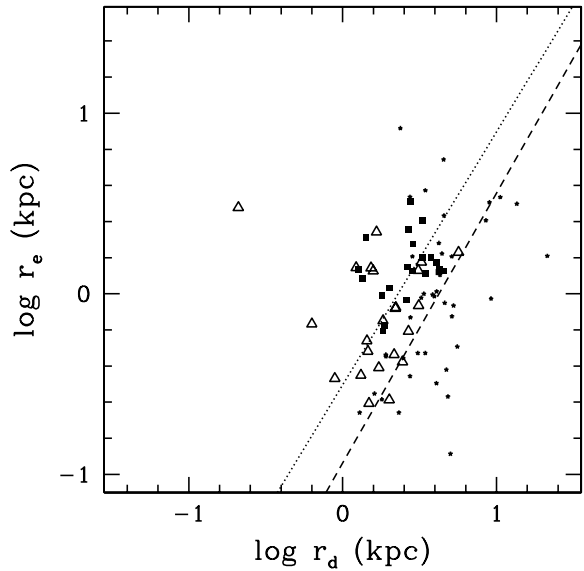,width=8.0cm,angle=0.}}
\caption{Dependence of the $\re - \rd$  relation on the  environment in
$K$ band  for the faint lenticulars. Late-type  spirals (as asterisks)
are overplotted for comparison. Other symbols are as in Figure 9. The dotted 
line is the best fit to the faint lenticulars (field as well as cluster) 
excluding the one obvious outlier while the dashed line is the best fit to
 late-type spirals. Note that these two lines have similar slopes but a different 
intercept.}
\label{f11}
\end{figure}

If lenticulars in clusters are indeed transformed spirals, it is likely
that they  preserve other signatures  of their earlier  existence. For
instance, if  they contained pseudobulges that  formed through secular
evolution, their  bulge and disk  sizes should be  correlated (Courteau
\etal  1996). In  Figure~\ref{f11},  we plot  the bulge  effective
radius  against  the  disk  scale  length for faint lenticulars (a similar 
plot for bright lenticulars may be found in Barway \etal (2007)). Faint  
field  and  cluster lenticulars  seems to both  follow the  same correlation,  
but inhabit different regions of  the plot. We also overplot data  for a 
sample of late-type  disk  galaxies  from  M\"{o}llenhoff \&  Heidt  (2001).   
The correlation  for  faint lenticulars  and  late  type  spirals has 
similar slopes ($\sim 1.40\pm0.14$  ) but different intercept which is
expected as late-type spirals have  larger disk scale length compared to
faint lenticulars. This is  a natural expectation if late-type spirals
are transforming  into faint lenticulars  due to interaction  with the
cluster medium as well as with other galaxies in the cluster.


\section{Summary}

The main results  of this paper  are: Several correlations
such  as the  Kormendy relation,  photometric plane  etc.  support the
hypothesis  that bright  and faint  lenticulars (in  cluster  or field
environments)  are fundamentally  different,  with different  formation
histories.  Bright  lenticulars  resemble  ellipticals and  bulges  of
early-type spirals suggesting that they may have formed like them - at
early epochs  via major mergers or rapid  collapse. Faint lenticulars,
on the other hand, have  properties consistent with them having formed
via  internal  secular  evolution  processes  (in the  field)  or  via
environment  influenced  secular  evolution  processes such  as  minor
mergers,   ram   pressure   stripping   and  galaxy   harassment   (in
clusters). Although the dominant differentiating parameter between the
two lenticular  classes is luminosity,  the environment also  seems to
play a  role in  determining the details  of lenticular  formation. In
particular, the  cluster environment seems  to induce a fading  of the
bulge  and the disk  and possible  transformation in  morphology from
spiral to lenticular, at least for the faint lenticular population.

If the  formation scenario of  bright and faint lenticulars  is indeed
completely different,  it should also manifest as  differences in star
formation history. These differences can be probed using a combination
of population synthesis models and  multiband bulge disk decompositions. 
We intend to carry
out such an analysis in a future project.

\section*{acknowledgements}

We thank the anonymous referee for insightful comments that improved the content and presentation of this paper. 
We thank A. G. Bedregal and A. Arag{\'o}n-Salamanca for providing us their
bulge-disk decomposition results in electronic form. We thank Somak Raychaudhury, 
Swara Ravindranath, S. K. Pandey and C. D. Ravikumar  for helpful
discussions. This research has made use of the NASA/IPAC Extragalactic
Database (NED), which is operated by the Jet Propulsion Laboratory,
California Institute of Technology, under contract with the National
Aeronautics and Space Administration. This publication makes use of data products from the Two Micron All Sky Survey, which is a joint project of the University of Massachusetts and the Infrared Processing and Analysis Center/California Institute of Technology, funded by the National Aeronautics and Space Administration and the National Science Foundation.


\begin{thebibliography}{99}
\bibitem[Abadi \etal (1999)]{1999MNRAS.308..947A} Abadi M.~G., Moore B., Bower R.~G., \ 1999, MNRAS, 308, 947
\bibitem[Barr \etal (2007)]{2007A&A...470..173B} Barr J. M., Bedregal A. G., Arag{\'o}n-Salamanca A. \etal \ 2007, A\&A, 470, 173
\bibitem[Barway \etal (2005)]{2005AJ....129..630B} Barway S., Mayya 
Y.~D., Kembhavi A.~K., \& Pandey S.~K., \ 2005, AJ, 129, 630
\bibitem[Barway \etal (2007)]{2007ApJ...661L..37B} Barway S., Kembhavi A., 
Wadadekar Y., Ravikumar C.~D., Mayya Y.~D., \  2007, ApJ, 661, L37
\bibitem[Bedregal \etal (2006)]{2006MNRAS..373..1125} Bedregal A. G., Arag{\'o}n-Salamanca A., \&
Merrifield M. R., \ 2006, MNRAS, 373, 1125 (BAM06)
\bibitem[Bekki (1998)]{1998ApJ...502L.133B} Bekki K., \ 1998, ApJ, 502, L133
\bibitem[Boselli \& Gavazzi (2006)]{2006PASP..118..517B} Boselli A., Gavazzi G., \ 2006, PASP, 118, 517
\bibitem[Bothun \& Gregg (1990)]{1990ApJ...350...73B} Bothun, G.~D., Gregg M.~D., \ 1990, ApJ, 350, 73
\bibitem[Buta \etal (2006)]{2006AJ....132.1859B} Buta, R., Laurikainen, E., Salo, H. \etal \ 2006, AJ, 132, 1859	
\bibitem[Courteau \etal (1996)]{1996ApJ...457L..73C} Courteau S., de Jong R. S., Broeils, A. H., \ 1996, ApJ, 457, L73
\bibitem[Cruz-Gonzalez \etal (1994)]{1994RMxAA..29..197C} Cruz-Gonzalez I., Carrasco  L., Ruiz E., Salas L. \etal \ 1994, RMxAA, 29, 197
\bibitem[Djorgovski \& Davis (1987)]{1987ApJ...313...59D} Djorgovski S., Davis M., \ 1987, ApJ, 313, 59
\bibitem[Dressler (1980)]{1980ApJ...236..351D} Dressler A., \ 1980, ApJ, 236, 351
\bibitem[Dressler (1987)]{1987ApJ...313..42D} Dressler A., \ 1987, ApJ, 313, 42
\bibitem[Dressler \etal (1997)]{1997ApJ...490..577D}  Dressler A., Oemler A. Jr. \etal \ 1997, ApJ, 490, 577
\bibitem[Gadotti \etal (2007)]{2007MNRAS.381..943G} Gadotti, D. A., Athanassoula, E., Carrasco, L., \etal \ 2007, MNRAS, 381, 943
\bibitem[Hubble(1936)]{1936RNeb..........H} Hubble E.~P., \ 1936, The Realm of the Nebulae, Yale University Press  
\bibitem[Jarrett \etal (2003)]{2003AJ....125..525J} Jarrett T.~H., Chester T. \etal \ 2003, AJ, 125, 525 
\bibitem[Khosroshahi \etal (2000a)]{2000ApJ...531L.103K} Khosroshahi H.~G., Wadadekar Y., Kembhavi A., \& Mobasher B., \ 2000a, ApJ, 531, L103 
\bibitem[Khosroshahi \etal (2000b)]{2000ApJ...533..162K} Khosroshahi H.~G., Wadadekar Y., \& Kembhavi A., \ 2000b, ApJ, 533, 162 
\bibitem[Kodama \& Smail (2001)]{2001MNRAS.326..637K} Kodama T., Smail I., \ 2001, MNRAS, 326, 637
\bibitem[Kormendy (1977)]{1977ApJ...218..333K} Kormendy J. \ 1977, ApJ, 218, 333
\bibitem[Kormendy \& Kennicutt(2004)]{2004ARA&A..42..603K} Kormendy J., \& Kennicutt R.~C. Jr., \ 2004, \araa, 42, 603 
\bibitem[Laurikainen et al.(2005)]{2005MNRAS.362.1319L} Laurikainen, E., Salo, H., \& Buta, R.\ 2005, MNRAS, 362, 1319 
\bibitem[Laurikainen et al.(2006)]{2006AJ....132.2634L} Laurikainen, E., Salo, H., Buta, R. \etal \ 2006, AJ, 132, 2634	
\bibitem[Lima Neto \etal (1999)]{1999MNRAS.309..481L} Lima Neto G.~B., Gerbal D., \& M{\'a}rquez I., \ 1999, MNRAS, 309, 481 
\bibitem[Metropolis \etal (1953)]{1953} Metropolis N., Rosenbluth, N., Rosenbluth, A., Teller, A., \& Teller, E. 1953, Journal of Chemical Physics, 21, 1087
\bibitem[Mobasher \etal (1999)]{1999MNRAS.304..225M} Mobasher B., Guzman R., Arag{\'o}n-Salamanca A., Zepf S. \ 1999, MNRAS, 304, 225
\bibitem[M\"{o}llenhoff \& Heidt (2001)]{2001A&A...368...16M} M\"{o}llenhoff C., Heidt J., \ 2001, A\&A, 368, 16
\bibitem[Moore \etal (1996)]{1996Natur.379..613M} Moore B., Katz  N., Lake  G., Dressler  A., \& Oemler  A., \ 1996, Nature, 379, 613
\bibitem[Moore \etal (1998)]{1998ApJ...495..139M} Moore B., Lake G., \& Katz N., \ 1998, ApJ, 495, 139
\bibitem[Moore \etal (1999)]{1999MNRAS.304..465M} Moore B., Lake G., Quinn T., \& Stadel J., \ 1999, MNRAS, 304, 465
\bibitem[Peletier \& Balcells (1996)]{1996AJ....111.2238P} Peletier R.~F., Balcells M., \ 1996, AJ, 111, 2238
\bibitem[Peletier \& Balcells (1997)]{1997NewAstro....1.349P} Peletier R.~F., \& Balcells M., \ 1997, New Astronomy, 1, 349
\bibitem[Peng \etal (2002)]{ 2002AJ....124..266P} Peng C. Y., Ho L. C., Impey C. D., Rix H., \ 2002, AJ, 124, 266
\bibitem[Poggianti \etal (1999)]{1999ApJ...518..576P} Poggianti B.~M. \etal \ 1999, ApJ, 518, 576
\bibitem[Simard \etal (2002)]{2002ApJS..142..1} Simard L. \etal \ 2002, ApJS, 142, 1 
\bibitem[Sersic (1968)]{1968AGA} S\'ersic J. L., \ 1968, Atlas de Galaxias Australes 
(Cordoba: Obs. Astron.) 
\bibitem[de Souza \etal (2004)]{2004ApJS..153..411D} de Souza R.~E., Gadotti D.~A., dos Anjos S., \ 2004, ApJS, 153, 411
\bibitem[van den Bergh (1994)]{1994AJ....107..153V} van den Bergh S., \ 1994, AJ, 107, 153
\bibitem[Wadadekar \etal (1999)]{1999AJ....117.1219W} Wadadekar Y., Robbason B., \& Kembhavi A., \ 1999, AJ, 117, 1219 

\end{thebibliography}
\end{document}